\documentclass[preprint]{aastex}

\begin{document}

\title{The Beginning of the End:  \textit{Hubble Space Telescope}
Images of Seyfert's Sextet}

\author{Christopher Palma, Stephanie G. Zonak, Sally D. Hunsberger,
Jane C. Charlton, Sarah C. Gallagher, Patrick R. Durrell}
\affil{Department of Astronomy \& Astrophysics, Pennsylvania
State University \\ Email:  cpalma@astro.psu.edu, szonak@astro.psu.edu,
sdh@astro.psu.edu, charlton@astro.psu.edu, gallsc@astro.psu.edu,
pdurrell@astro.psu.edu}
\authoraddr{525 Davey Laboratory, University Park, PA 16802}

\and

\author{Jayanne English}
\affil{University of Manitoba\\ Email: 
Jayanne\_English@umanitoba.ca}
\authoraddr{}

\begin{abstract}

\textit{Hubble Space Telescope} Wide Field Planetary Camera 2 images of
Hickson Compact Group 79, Seyfert's Sextet, are presented.  Both point
sources and extended sources detected on the three WF chips were
photometered in four filters:  F336W, F439W, F555W, and F814W.  Unlike
other HCGs that have been imaged with \textit{HST}, there do not appear
to be any candidate young star clusters among the detected point
sources.  The majority of the point sources that may be star clusters
associated with the Sextet have red colors consistent with stellar
populations older than 1 Gyr.  A similar conclusion is drawn with
regard to the extended sources.  The majority of these appear to be
background galaxies, but a few candidate dwarf galaxies are identified
as potentially associated with Seyfert's Sextet.  However, no blue,
star forming objects similar to the tidal dwarf galaxy candidates
identified in other HCGs are found among the extended objects
identified in this study.  A redshift for one dwarf galaxy candidate
was measured from a spectrum obtained with the Hobby-Eberly Telescope,
and this object was found to have a redshift similar to NGC6027e, the
discordant spiral formerly identified as a member of this compact
group.  The \textit{HST} observations presented here and previous radio
observations of the neutral gas content of this group suggest that the
interactions that have taken place in the Sextet only redistributed the
stars from the member galaxies within the group.  We speculate that future
interactions may be strong enough to strip the gas from NGC6027d and
trigger star cluster formation.

\end{abstract}


\section{Introduction}

Compact groups of galaxies are found at the extreme end of the
distribution of galaxy surface densities; by definition, compact groups
consist of four or more isolated galaxies found within a small area on
the sky.  Using the \citet{hickson82} compact group selection criteria,
the surface densities of galaxies in compact groups are similar to or
larger than those found in the centers of massive galaxy clusters.   A
radial velocity survey of the 100 Hickson compact groups \citep[HCGs;
][]{hickson92} has verified that the majority of these HCGs are
physical associations of 3 or more galaxies with a median velocity
dispersion of 200 km sec$^{-1}$.  One expects that in such dense groups
with small velocity dispersions that interactions and mergers among the
members are inevitable.

As expected, the morphologies of the galaxies in compact groups often
show evidence of tidal interaction.  Galaxy mergers are complicated
phenomena, and it is difficult to disentangle the past histories of the
merging galaxies, even in isolated, merging pairs.  In compact groups
there is often evidence for multiple interactions
\citep[e.g.,][]{hickson82}, so unravelling the history of these systems
is even more of a challenge.  Fortunately, star formation is a useful
tool for dating some of the discrete events in the interaction history
of a galaxy merger.

Relying on \textit{Hubble Space Telescope (HST)} images, a growing number of
studies have found evidence for compact star cluster formation in
systems of interacting galaxies or merger remnants \citep{holtzman92,
whitmore93, rwo95, schweizer96, miller97, zepf99, johnson00, sq01}.
Using stellar population synthesis models, the ages of the observed
young star clusters can be estimated from their photometric colors.
In cases where discrete populations of young clusters are found in different
regions of an interaction or merger remnant, variations among the ages of
the clusters can be used to date events in the merger history.  For example,
\citet{whitmore99} identified four distinct populations of young star clusters
in the ``Antennae'' (NGC 4039/39), and used the ages of these populations to
infer some of the evolutionary history of this system.  \citet{sq01} have
identified a number of compact clusters in HCG 92, Stephan's Quintet.  This
system is more complicated than the Antennae, but the star cluster ages
have been used to identify distinct epochs of cluster formation and to
date some of the interaction events.

Star formation in interacting galaxies is not necessarily limited to
compact cluster formation; there is also observational evidence for the
formation of extended, dwarf galaxy-sized objects in the tidal debris
of a number of systems \citep[e.g.,][]{mirabel91, mirabel92, hibbard94,
deeg98, duc98, duc00, MdO01}.  \citet{sdh96, sdh98} have performed a
statistical analysis of a large sample of HCGs and found evidence that
tidal dwarf galaxy formation in compact groups may be common.
Numerical simulations suggest that dwarf galaxy formation in tidal
tails is possible \citep{barnes92, elmegreen93}; however, it is unclear
whether the observationally identified ``tidal dwarf galaxies'' (TDGs)
will evolve into the bound entities seen in the simulations.  Recently,
a few groups have presented dynamical analyses of several TDGs
\citep{duc00, MdO01, hibbard01} in order to assess the likelihood that
these objects will remain bound, but the results remain ambiguous.

\textit{HST} imaging of HCGs is useful for addressing both the
interaction history of a compact group as well as the presence and
nature of any putative TDGs.  In this paper, we present Wide Field
Planetary Camera 2 (WFPC2) observations of Hickson Compact Group 79,
Seyfert's Sextet, taken in the $U$, $B$, $V$, and $I$ (F336W, F439W,
F555W, and F814W) filters.  In Sections 2.2--2.4, the compact clusters
in the group are identified and their ages estimated from stellar
population models.  In Sections 2.5--2.6, we present the extended
objects in the field and discuss their nature.  Section 3 discusses a
spectroscopic follow-up observation of one candidate dwarf galaxy
associated with the Sextet.  Finally, in Section 4, we compare the
population of compact clusters and dwarf galaxies in Seyfert's Sextet
to those found in the recent \textit{HST} studies of HCG 92 \citep[Stephan's
Quintet;][]{sq01} and HCG 31 \citep{johnson00}.

\section{Observations of Seyfert's Sextet}

\begin{figure}

\plotone{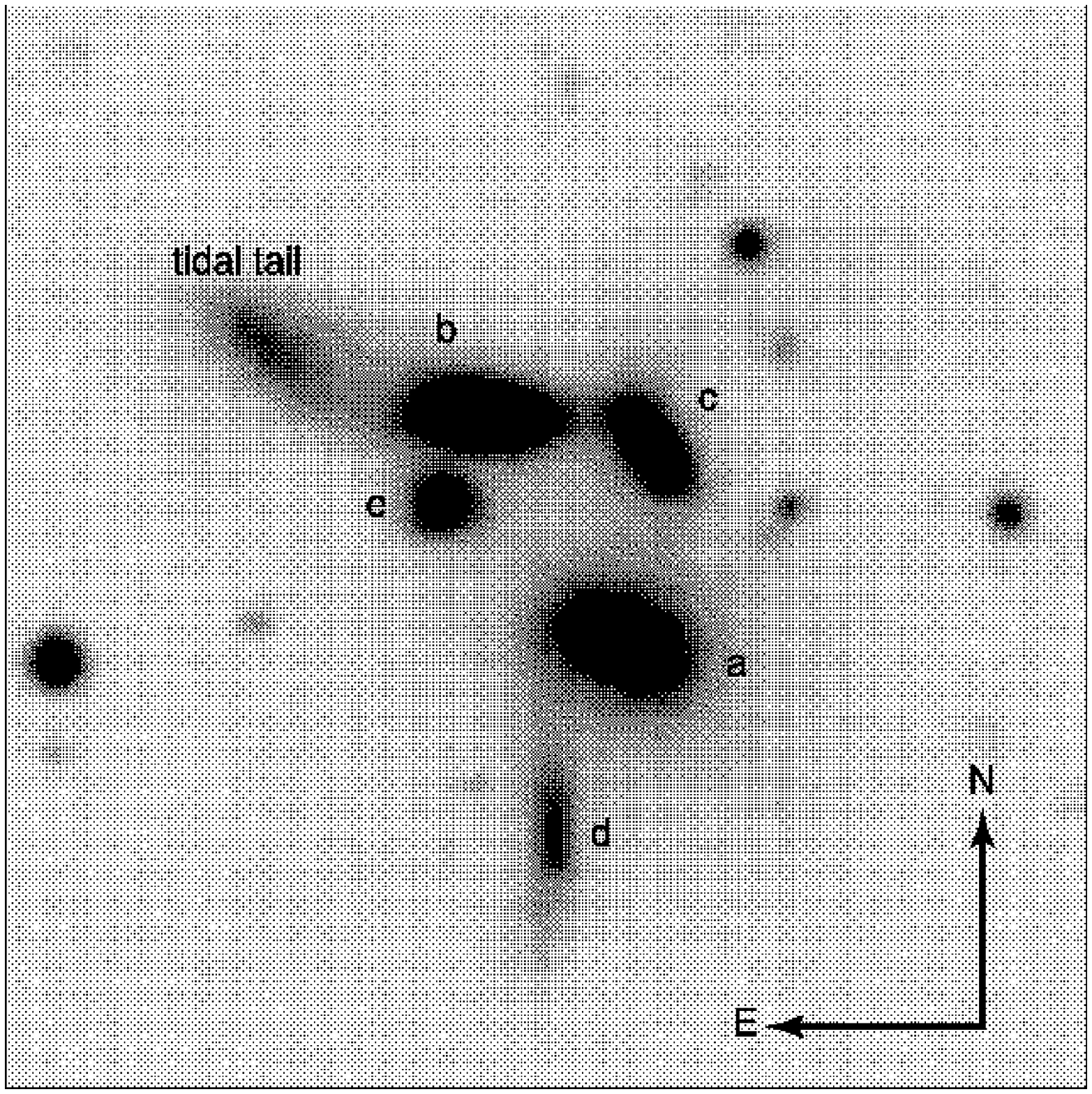}

\caption{An image of Seyfert's Sextet taken from the second generation
Digitized Sky Survey (the red plate).   This image is 5\arcmin\ on a
side and north is up, east to the left. The individual galaxies,
NGC6027a-e, are labeled using the notation of \citet{hickson82}.  The
bright tidal tail associated with NGC6027b is also labeled.  There
is another, fainter tidal tail associated with NGC6027c, but it is 
not seen in this image.}

\label{dssimage}

\end{figure}

\citet{cks48, cks51} was the first to identify and study this
``unusually densely crowded group of six galaxies'' from photographic
plates taken with the Harvard Schmidt telescope.  While the second
article refers to six galaxies and does name the group a sextet,
\citet{cks51} mentions that Baade's observations of the group show that
the object labeled ``NGC6027e''\footnote{Note that Seyfert's
designations for the individual galaxies, NGC6027 and NGC6027a-e, are
not the same as the commonly adopted, modern designations for the
galaxies.  Throughout this article, we use the designations adopted later by
\citet{hickson82}.  The galaxies in the Sextet are labeled in the image
presented as Figure \ref{dssimage}.} is ``not a separate galaxy but a
tidally distorted part of NGC6027.''  \citet{cks51} goes on to say that
``If this is actually the case, we have a very extraordinary filament
which not only has a perceptible condensation but a filament which is
nearly twice as large as the parent spiral.''  Based on the
morphological differences between the two late-type spirals and the
tidally distorted early-type spirals, \citet{cks51} considered two
possibilities:  (1) the Sextet is truly comprised of three physically
associated early-type galaxies, two background spirals, and one bright
tidal tail, or (2) it is a physically associated group of six
galaxies.

Seyfert's Sextet fits the criteria for inclusion in the Hickson compact
group catalogue \citep{hickson82}.  It is the most compact of the HCGs
\citep{hickson97}, with a median galaxy separation of $9.1 h_{75}^{-1}$ kpc
(we adopt $H_0 = 75$ km s$^{-1}$ Mpc$^{-1}$ throughout this paper).
The group is now known to consist of four galaxies with accordant
redshifts, one bright tidal tail, and a fifth galaxy at a redshift much
larger than the group median redshift \citep{hickson92}.  Using the
mean redshift of the accordant galaxies, $z=0.0145$ \citep{hickson92},
the distance to the Sextet is 57.5$h^{-1}$ Mpc and the distance modulus
is 33.8 magnitudes.  A summary of the global properties of Seyfert's
Sextet is given in Table \ref{SSparams}.  In Figure \ref{dssimage}, we
present an image of Seyfert's Sextet from the second generation red
Digitized Sky Survey\footnote{The Digitized Sky Survey was produced at
the Space Telescope Science Institute under U.S.\ Government grant NAG
W-2166.  The images of these surveys are based on photographic data
obtained using the Oschin Schmidt Telescope on Palomar Mountain and the
UK Schmidt Telescope.  The plates were processed into the present
compressed digital form with the permission of these institutions.  The
Second Palomar Observatory Sky Survey (POSS-II) was made by the
California Institute of Technology with funds from the National Science
Foundation, the National Aeronautics and Space Administration, the
National Geographic Society, the Sloan Foundation, the Samuel Oschin
Foundation, and the Eastman Kodak Corporation.} that includes labels
for the individual member galaxies using the scheme of
\citet{hickson82}.

\subsection{\textit{HST} Observations}

The WFPC2 instrument on the \textit{Hubble Space Telescope} was used to
observe Seyfert's Sextet in 2000 June (program ID 8717, Hunsberger
PI).  The observations were planned in such a way as to place the
Sextet galaxies on the three Wide Field chips.  Because of the
placement of the galaxies of interest on the three WF chips, and since
additional read noise in the PC data raises the detection threshold for
point sources on this chip compared to the WF chips, the Planetary
Camera data were not included in this study.  The total exposure time
in each filter was broken up into four cosmic ray-split exposures.  One
dithering offset of $\sim$2.5 pixels in the $x$ and $y$ directions was
introduced so that two of the four subexposures are at one dither
location and the other two subexposures are at another.  The exposure
times were $4 \times 500$ seconds in F814W and F555W and $4 \times
1300$ seconds in F439W and F336W.  The gain was set to 7 $e^-$/ADU
during these exposures.

In order to use the latest \textit{HST} calibration, the data were
recalibrated with the ``best'' reference files available from the
\textit{HST} archive in 2002 January.   Known hot pixels were removed
from the images using the IRAF\footnote{IRAF is distributed by the
National Optical Astronomy Observatories, which is operated by the
Association of Universities for Research in Astronomy, Inc., under
cooperative agreement with the National Science Foundation.} STSDAS
task WARMPIX.  After hot pixel removal, the two images at each dither
position in each filter were averaged together with GCOMBINE and then
further cleaned of cosmic rays with the IRAF task COSMICRAYS.
Identification and photometry of point sources (see \S 2.2) were done
individually on each of the images of the two dither locations and the
results subsequently averaged.  The choice to analyze each dither
location separately was made because the detection of extremely faint
sources and the study of the sizes of these objects are not the focus
of this paper. Using the IRAF tasks GEOMAP and GEOTRAN to solve for the
transformation between the two dither locations, a final, combined
image in each filter that retains the original pixel scale was
created.  Identification and photometry of the extended sources in the
field were done on the final, combined image (see \S 2.5).

Also, a color image of Seyfert's Sextet (Figure \ref{primage}) was created
by combining the black and white data collected through the 4 filters.
The intensities were first clipped and scaled to bring out faint
emission. (KARMA's Kview task was employed for this step but the
remainder of the manipulations were accomplished using GIMP.) Each
filter's scaled image was then assigned a color that approximated the
peak wavelength of that filter: F336W was colored reddish-violet,
F439W blue, F555W yellowish-green, and F814W red. These colour images
were combined using the ``screen'' algorithm in GIMP. The contrast of
the combined image was adjusted to produce a neutral, dark background
and emphasize particular features.  It is the inclusion of the
ultraviolet filter that causes the disk of the central spiral galaxy,
as well as the central regions of the edge-on spiral, to be rendered
more pink than they would be if strictly optical filters were used.

\subsection{Point Source Detection}

The method used for point source detection follows very closely the
technique used by \citet{sq01}.  A brief summary of this method is
presented here:  Each dither location in each filter was initially
searched for point sources using the DAOFIND task within IRAF's
implementation of DAOPHOT \citep{daophot}.  The detection threshold was
set very low in order to find potential sources in regions of higher
than average background (e.g., in the faint tidal tail).  In order to
remove some spurious detections, only those objects detected at the
same location in both dither images were retained.  Initial aperture
photometry was performed on this large number of objects (thousands per
WF chip), and all objects with $S/N < 3.0$ at either or both dither
locations were thrown out.  

Although this process serves to eliminate most of the spurious sources,
a number of objects remain in the catalogue that can be eliminated as
potential compact star clusters associated with Seyfert's Sextet.
Adopting the mean group redshift as a distance measure, the Sextet is
$\sim$58 $h^{-1}_{75}$ Mpc from Earth, and at this distance, one pixel
is $\sim$27 $h^{-1}_{75}$ pc.   Typical Galactic globular clusters have
$r<<27$pc \citep{mwgc}, and even large, young clusters seen in the
Antennae have effective radii (the radius enclosing half of the light,
seen in projection) of $r_{eff} < 15$ pc, which leads us to
expect that compact clusters in the Sextet will be unresolved.  Thus,
we can clean the catalogue further by eliminating all 
resolved sources.  

Resolved sources were removed from the catalogue using the same set of
tests used by \citet{sq01} \citep[which are derived from those
in][]{miller97}.  Specifically, at each dither location, every
potential point source was photometered with an aperture radius of 0.5
pixels and a second time with an aperture radius of 3.0 pixels.  The
difference in the $V$ magnitude between the two apertures, $\Delta
V_{0.5-3.0}$ was calculated, and all objects with $\Delta V_{0.5-3.0} <
0.5$ (likely cosmic rays or other hot pixels) or $\Delta V_{0.5-3.0} >
2.5$ (resolved sources) in either pointing were flagged.  Also, using
the IRAF task RADPROF, fifth order cubic splines were fit to the radial
profiles of each potential point source.  In all four filters, the mean
FWHM for the likely point sources is $1.0 - 1.1$ pixels, however the
peak of the four distributions are near 0.9 pixels, with a non-Gaussian tail
to higher values.  To separate real point sources from resolved
sources, we chose to flag those objects with FWHM $> 2.0$ pixels,
also.  Each object was thus tested four times:  once with the $\Delta
V_{0.5-3.0}$ test and once with the FWHM test at each dither location.
Any object with two or more flags was removed from the catalogue of
point sources.

The final catalogues did contain some sources that passed all
algorithmic tests, but could still be easily ruled out as potential
compact star clusters.  These included bright, foreground, Galactic
stars, the unresolved nuclei of a few background galaxies, and
\ion{H}{2} regions in the disk of the background galaxy in the Sextet
(NGC 6027e).  The foreground stars and galactic nuclei are easily
distinguishable from the candidate star clusters.  Although the
\ion{H}{2} regions in NGC 6027e are blue point sources that might be
mistaken for star cluster candidates, they are all aligned 
with the prominent spiral arms in this disk galaxy, so we are confident
that they are unlikely to be point sources associated with Seyfert's
Sextet.  We removed by hand these three types of contaminating objects
from the catalogue.  In the end, we were left with a total of 188 star
cluster candidates (see Figure \ref{fchart1} for a finding chart)
detected in at least one filter.  Very few objects were detected in the
two blue filters, F439W and F336W; we discuss this further in \S2.4.

\subsubsection{Sub-pixel Dithering}

The exposures of Seyfert's Sextet were taken at two dither locations
and can be combined in order to enhance the angular resolution using
the ``DRIZZLE'' software developed by \citet{driz}.  Since the cluster
population of Seyfert's Sextet should be unresolved even in drizzled
images, we chose to apply our cluster detection algorithm to the raw,
undrizzled images rather than to drizzled images.  However, we did
combine the F439W and F555W images of the WF2 chip with the drizzle
software in order to test the differences in cluster detection on
drizzled and undrizzled images.  Drizzle was run with the parameters
$pixfrac=1.0$ and $scale=0.5$, which is equivalent to the ``shift and
add'' technique, and the resulting images have a pixel scale twice as
fine as that of the input images.

We used an identical cluster detection algorithm to the one described
above:  The objects detected by DAOFIND in the F439W and F555W drizzled
and undrizzled images were pruned first by signal-to-noise and then
were pruned further using the FWHM and $\Delta V_{0.5-3.0}$ (the
aperture sizes were doubled to 1.0 pix and 6.0 pix for the drizzled
image) tests discussed in detail in the previous section.  The most
significant cut is the signal to noise cut; since the threshold used in
DAOFIND is so low, many of the objects detected have $S/N < 2.5$.
Furthermore, there are significant (at the $\sim10$\% level) differences
in $S/N$ between the objects in the drizzled image and in the undrizzled
image.  This is especially true of objects found in regions where the
background is variable, such as within the bright isophotes
of NGC6027a and NGC6027d.  After applying our signal to noise criterion,
a number of (likely spurious) objects within the bright isophotes of these
two galaxies remain in the catalogue created from the undrizzled image.
However, in comparison, many of these get removed from the catalogue created
from the drizzled image.

The second criterion used to flag objects for removal from our point
source catalogues are the ``structure'' parameters -- FWHM and the
concentration or $\Delta V_{0.5-3.0}$.  What we find is that objects
that have $0.5 < \Delta V_{0.5-3.0} < 2.5$ lie in a narrow range of
FWHM, indicating that they are likely to be point sources.  Almost
every object within this window of concentration parameter have FWHM $<
2.0$ pixels in the undrizzled image and FWHM $< 4.0$ pixels in the
drizzled image.  For objects with $\Delta V_{0.5-3.0} > 2.5$, the range
of FWHM found for objects in both the drizzled and undrizzled images is
much larger, extending from small values typical of point sources all
the way to FWHM $\sim$15 pixels or so.  Thus, it appears that  $\Delta
V_{0.5-3.0} \sim 2.5$ signals the transition from point sources to
resolved sources. 

Our detailed comparison between drizzled and undrizzled images of the
WF2 chip in filters F439W and F555W shows that in the end the point
sources identified are nearly identical.  The largest difference found
is in $S/N$, and results in more spurious objects being identified in
the disk of NGC6027d and near the dust lane in NGC6027a in the
undrizzled image than in the drizzled image.  Although the pixel scale
is different by a factor of two between the drizzled and undrizzled
images, the results for the structure parameter are nearly identical
for point source candidates in both images:  Those objects with $0.5 <
\Delta V_{0.5-3.0} < 2.5$ have FWHM values expected for point sources
in both the drizzled and undrizzled images.  Those objects with $\Delta
V_{0.5-3.0} > 2.5$ have FWHM values typical of resolved objects.  Thus,
we conclude that for the dataset presented here, our point source
detection algorithm is not significantly improved by the application of
the drizzling technique.

We note that a recent study of the star cluster candidates in NGC 3610
by \citet{whitmore02} supports this conclusion.  In a previous study of
NGC 3610, \citet{whitmore97} used undithered images to identify point
sources using an algorithm almost identical to the one used here.  
In the more recent study, \citet{whitmore02} drizzled together deeper
exposures of NGC 3610.  Even though the $S/N$ and pixel scale were 
improved in the more recent images, the results of the new search were
similar to the results of the original study.  They found that 22 of 23 
cluster candidates identified from the earlier images were recovered
in the new images.  Also, 11 new, previously unidentified clusters were
discovered in the higher resolution images.  However, the majority of these
are within the disk of NGC 3610, and the authors attribute their detection
to their ability to identify clusters in regions of higher background, which
is likely because of the combination of improved $S/N$ and pixel scale,
not just because of the improved pixel scale.

\subsection{Aperture Photometry of Point Sources}

We performed circular aperture photometry of all point sources detected
using the method described previously.  For consistency with
\citet{sq01}, the photometry of each source was calculated using
similar techniques.  The technique used and the few differences between
this study and \citet{sq01} are described here.

Each object was measured with the IRAF APPHOT package using a circular
aperture of 2 pixel radius.  The sky level for each object was
determined by taking the median value in an annulus of 7 pixel inner
radius and 10 pixel outer radius centered on the object.  A correction
for charge transfer efficiency (CTE) was calculated for each object
using the formulation given in \citet{dolphin00}.  Photometry was
measured at both dither positions for all point sources, the CTE
correction was applied, and then the results for the two positions were
averaged.

In order to transform the instrumental magnitudes to the standard
VEGAMAG system, the zero points from \citet{dolphin00}\footnote{The
functional form and coefficients for the CTE correction as well as the
photometric zero points were taken from the website update to the
published values, which is found at the following URL:
www.noao.edu/staff/dolphin/wfpc2\_calib/.} were adopted.  These zero
points were calculated using the standard $0\farcs5$ aperture, and
therefore, aperture corrections need to be applied to our data to
account for the different aperture size used.  

Aperture corrections were calculated by photometering bright, isolated
point sources in our 2 pixel aperture and in the 5.15 pixel, $0\farcs5$
aperture.  The difference in magnitudes in these two measurements was
calculated for each point source, and the mean adopted as the aperture
correction.  There were not enough point sources to solve for the
aperture correction for each WF chip independently, so we adopt the
same aperture correction for each chip.  Table \ref{apcor} lists the
adopted values for the aperture correction in each filter, the number
of objects used to derive the correction, and the standard deviation of
the measurements.

The value for the Galactic extinction towards Seyfert's Sextet was
derived from the \citet{schlegel} dust maps.  The value returned was
$E(B-V) = 0.055$, which is adopted as the foreground reddening towards
all objects in the WFPC2 images of the Sextet.  In order to determine
the reddening corrections for the specific \textit{HST} filters
employed in this study, a synthesized instantaneous-burst star cluster
model spectrum \citep[the ``BC95'' models;][]{bc93, charlot96} with an
age of 10 Gyr (see \S2.4) was reddened using a \citet{cardelli}
extinction law.  Using the STSDAS SYNPHOT package, this reddened
spectrum was convolved with the \textit{HST} filter functions and the
following corrections were calculated: $A_{F336W} = 0.270$, $A_{F439W}
= 0.228$, $A_{F555W} = 0.171$, and $A_{F814W} = 0.101$.  All of the
point source photometry presented here has been corrected using these
estimates for the Galactic reddening corrections.

\subsection{Results of the Point Source Photometry}

\begin{figure}

\plotone{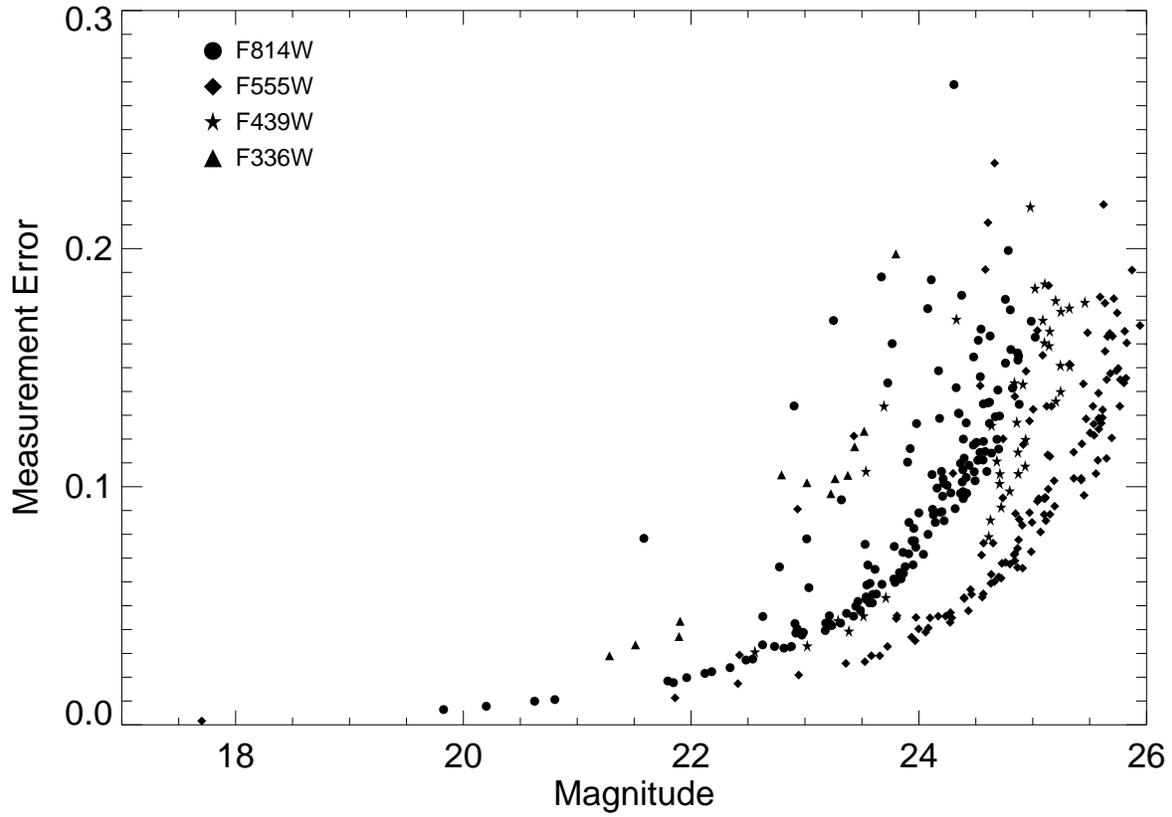}

\caption{A plot of photometric error (in magnitudes) for the point
sources identified in the three Wide Field Camera chips.  This error
only reflects measurement error, and does not include the contribution
from the transformation to the standard system or the CTE correction.
Different symbols are used for each filter:  F814W data are the filled
circles, F555W diamonds, F439W stars, and F336W triangles.}

\label{photerr}
\end{figure}

\begin{figure}

\plotone{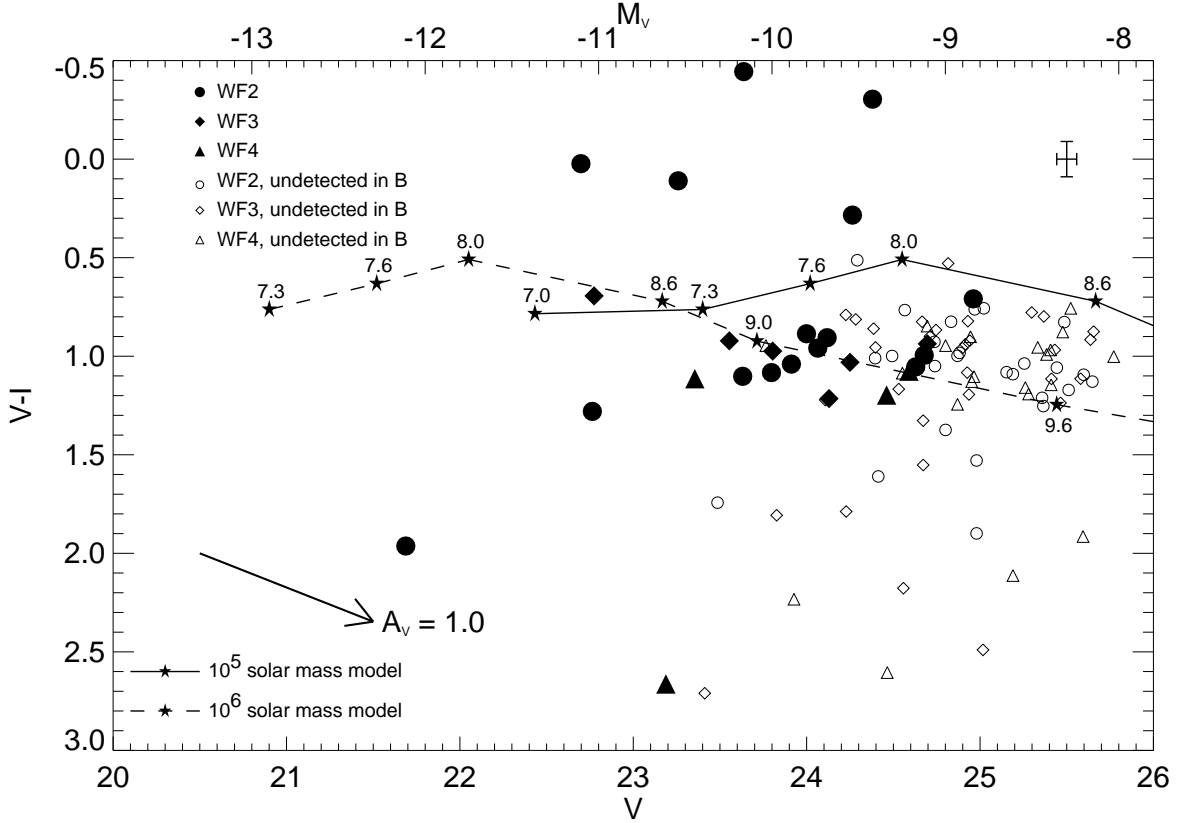}

\caption{Color-magnitude diagram for all point sources detected in $V$
(F555W) and $I$ (F814W).  All point sources detected in $B$, $V$, and
$I$ are plotted with filled symbols:  circles for objects on Wide Field
chip 2, diamonds for Wide Field chip 3, and triangles for Wide Field
chip 4.  Those objects detected in $V$ and $I$ but not detected in $B$
are plotted as open symbols with the same symbol shapes reflecting chip
location.  A reddening vector is included for reference (lower left) as
is the mean error in the photometry for the filled symbols (upper
right).  All of the photometry has been corrected for foreground
Galactic reddening assuming $E(B-V) = 0.055$.  The absolute magnitude
scale is given assuming a distance to the Sextet of 57.5 $h_{75}^{-1}$
Mpc.  The two sets of solid stars plotted indicate the colors and
magnitudes for an instantaneous--burst stellar population synthesis
model \citep{charlot96} of various ages; the points connected with the
solid line and the dashed line are for models with total masses of
$10^5 M_{\sun}$ and $10^6 M_{\sun}$, respectively.  The model points
are labeled with the logarithm of the age of the model, in years.  The
colors of the model stellar population are independent of the cluster
mass, however changes in cluster mass will shift the model points by
one magnitude for each factor of 2.5 change in the mass.}

\label{CMD}
\end{figure}

\begin{figure}

\plotone{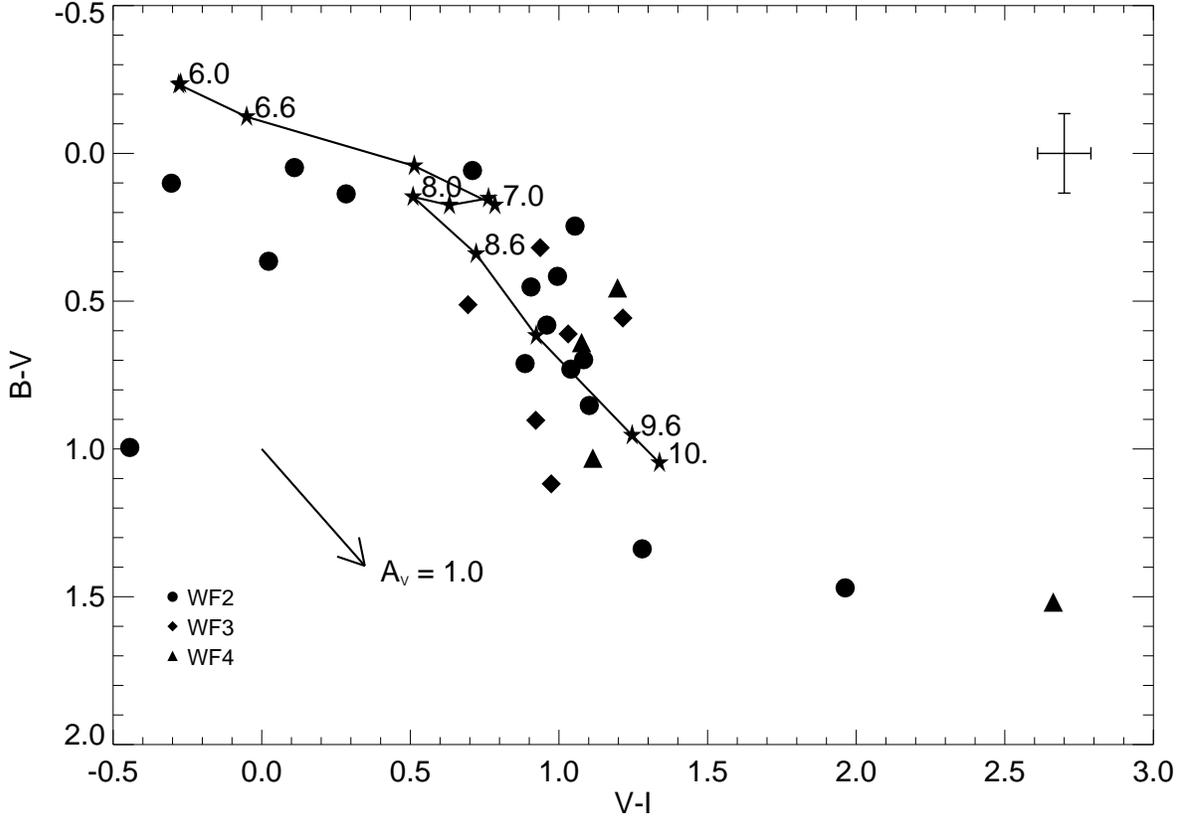}

\caption{Color-color diagram for all point sources detected in $B$
(F439W), $V$ (F555W), and $I$ (F814W).  All of the photometry has been
corrected for foreground Galactic reddening assuming $E(B-V) = 0.055$.
Symbols are as in Figure \ref{CMD}.  A reddening vector and error bars
representing the mean error for the objects plotted are included for
reference.  The solid stars indicate the colors for an
instantaneous--burst stellar population synthesis model
\citep{charlot96} of various ages.  Several of the points are labeled
with the logarithm of the age of the model, in years.}

\label{CCD}
\end{figure}

Figure \ref{photerr} presents the calibrated photometry and associated
measurement error for the 188 point sources identified in the three
Wide Field chips.  Among the 188 point sources detected, 106 were
simultaneously detected in both $V$ and $I$, 26 in $B$, $V$, and $I$,
and only 5 were detected in all four filters.  Figures \ref{CMD} and
\ref{CCD} are the $V - I$ vs. $V$ color--magnitude diagram (CMD) for
the 106 sources detected in both filters and the $B - V$ vs. $V - I$
two--color diagram (2CD) for the 26 sources detected in all three
filters.

Although a correction for Galactic reddening was applied to the point
source photometry, differential reddening due to the dust content of
the compact group itself is also likely to affect the photometry of the
point sources.  For reference, a reddening vector equivalent to one
magnitude of extinction in the F555W filter is plotted in Figures
\ref{CMD} and \ref{CCD}.  Differential reddening is likely to be
affecting many of the sources to some degree.  However, because at least
some of the point sources are found far from the outer isophotes of the
giant galaxies, it is unlikely that there is a large, systematic shift
in the colors and magnitudes of all of the sources.

If we assume that the differential reddening in the group is not
severe, then one conclusion that can immediately be drawn from the CMD
and the 2CD is that there are very few bright, blue point sources
detected.  Our choice of exposure times for the two blue filters was
guided by observations of blue point sources in other studies of
mergers \citep[e.g., the Antennae,][]{whitmore99} and compact groups
\citep[e.g., HCG92,][]{sq01}.  Given our exposure times, the resulting
limiting magnitude in $B$ is brighter than our $V$ limit (the
magnitudes where the mean photometric error for point sources reaches
0.1 mag are 24.6, 25.0, and 24.3 for $B$, $V$, and $I$, respectively),
however, the limits are such that we should have detected any objects
with $B - V \lesssim 1.0$ and $V \lesssim 23.5$ in both filters.  This
observational selection effect does render us insensitive in $B$ to
faint point sources with $B - V \gtrsim 1.0$, however, the candidate
young star clusters identified in other systems are all blueward of
this limit.

In Seyfert's Sextet, we find only a handful of objects are detected
blueward of $V - I = 0.7$.  The majority of the point sources detected
have $0.7 < V-I < 1.3$ (which are likely to have $B - V \sim 0.7 -
1.0$), but most went undetected in $B$.  This lack of bright, blue
point sources in the Sextet is somewhat unexpected since these objects
are found in abundance in similar systems.  For example, in
\citet{sq01}, the peak in the color histograms for the identified point
sources in Stephan's Quintet (HCG92) are found at $B - V \sim 0.1$ and
$V - I \sim 0.4$.  Also, \citet{johnson00} find a number of bright
point sources in \textit{HST} images of HCG31 that have $-1 < V - I <
0.4$ (the extremely blue objects in this sample are likely to have a
large contribution to their flux from nebular line emission).

The 2CD for Seyfert's Sextet point sources (Figure \ref{CCD}) also
includes the evolutionary tracks derived from the BC95 \citep{bc93,
charlot96} solar metallicity instantaneous--burst stellar population
models.  In the CMD, the colors and magnitudes of these same BC95
models are included, where the magnitudes have been normalized for a
cluster with masses of $10^5 M_{\sun}$ and $10^6 M_{\sun}$.  There is a
significant amount of scatter in the data for the star cluster
candidates, however, the majority of the objects detected have colors
in the 2CD similar to the expectations for clusters of 1 Gyr or older.  In the
CMD, the track for the $10^6 M_{\sun}$ clusters fits the data better
than that of the lower mass model, and also indicates an age for the
clusters in excess of 1 Gyr. Unlike the two previous studies of
star cluster formation in compact groups \citep{johnson00, sq01}, there
does not appear to be any significant population of candidate young
star clusters with ages less than 1 Gyr.

\begin{figure}

\epsscale{0.67}
\plotone{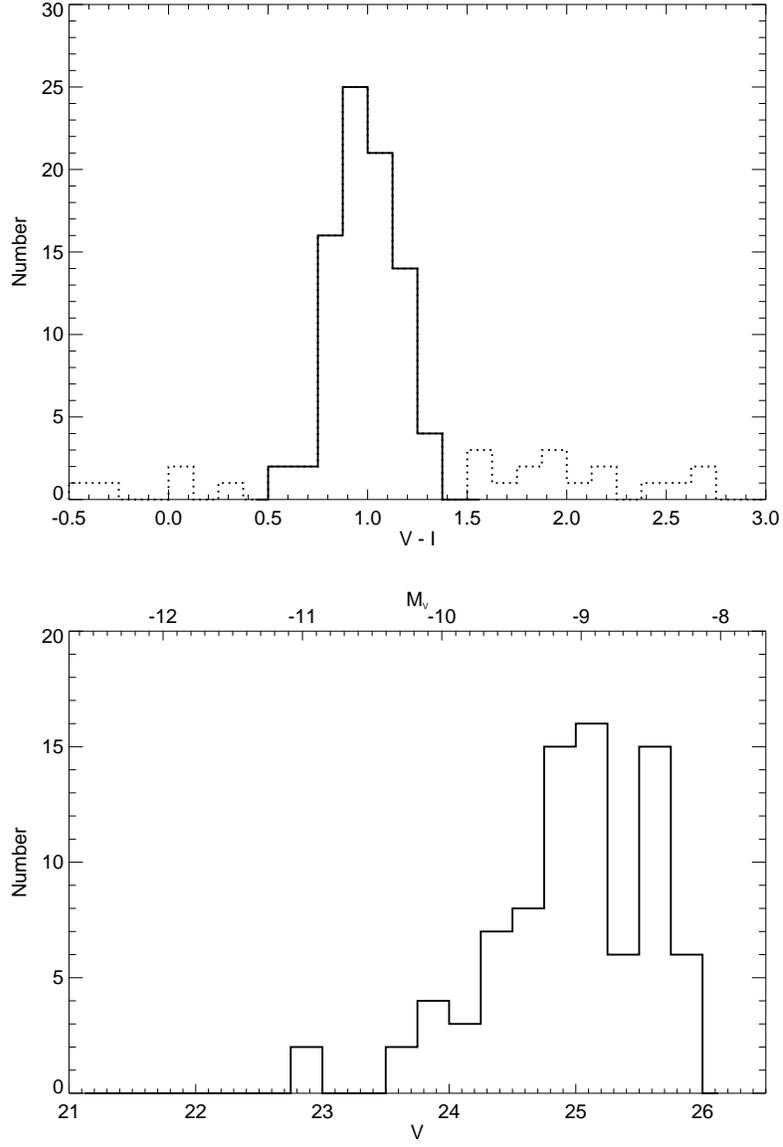}

\caption{Histograms of the $V-I$ colors for 106 point sources detected
on the WF chips (top panel) and $V$ magnitudes for the 84 likely
cluster candidates in the restricted color range $0.5 < V-I < 1.5$
(bottom panel).   The bins of $V-I$ color represented by the dotted
line in the top panel represent those objects that were excluded from
the $V$ magnitude histogram in the bottom panel.}

\label{hists}
\end{figure}

In Figure \ref{hists}, we present histograms of the $V-I$ colors for
the 106 point sources detected in both $V$ and $I$ (top panel) and the
$V$ magnitudes for the 84 likely cluster candidates in the restricted
color range $0.5 < V-I < 1.5$ (bottom panel).  The color and magnitude
distributions presented in this Figure are consistent with our
interpretation of an older population of globular clusters.  The mean
color of the objects in the Sextet with $0.5 < V - I < 1.5$ is 0.99,
which is similar to the mean colors of the globular cluster populations
in \textit{HST} observations of ellipticals and S0s found by
\citet{kundu01a, kundu01b}.   The histogram of the $V$ magnitudes of
the likely cluster candidates identified in the Sextet is consistent
with the luminosity function of the globular cluster populations seen
in the deeper photometry of \citet{kundu01b}; some of the single
elliptical galaxies in that study contain more clusters brighter than
$V = -9$ than are found in the entire compact group studied here.
However, since the observations presented in this paper were not
designed to study the fainter old population of globular clusters in
this compact group, it is likely that our census of the old clusters in
the group is incomplete.  Given the likely incompleteness, though, the
data on the Sextet cluster candidates presented in Figure \ref{hists}
are consistent with the standard (old) globular cluster luminosity
function \citep{harris01}, not just the \textit{HST} data of
\citet{kundu01b}.

Blue globular clusters are expected to be as bright or brighter than
the older, red globular clusters (unless they are of much lower mass);
for example, many of the blue clusters identified in NGC3256
\citep{zepf99} have $M_I < -10$, which is above our detection threshold.
Thus, Figures \ref{CMD} and \ref{hists} should represent accurately the
population of clusters in the Sextet; the population is truly dominated
by red clusters.  However, among the sources that were detected in $B$,
$V$, and $I$, there are a few blue point sources observed, those on WF2
that have $V - I < 0.5$.  Four of the five blue point sources are
located within the disk of NGC6027d, and the fifth is just south of the
disk (it is the southernmost point source visible in Figure
\ref{fchart1}).  This region appears to be the only site that includes
young, star-forming regions.

\subsection{Extended Source Detection and Photometry}

\begin{figure}

\epsscale{1.00}
\plotone{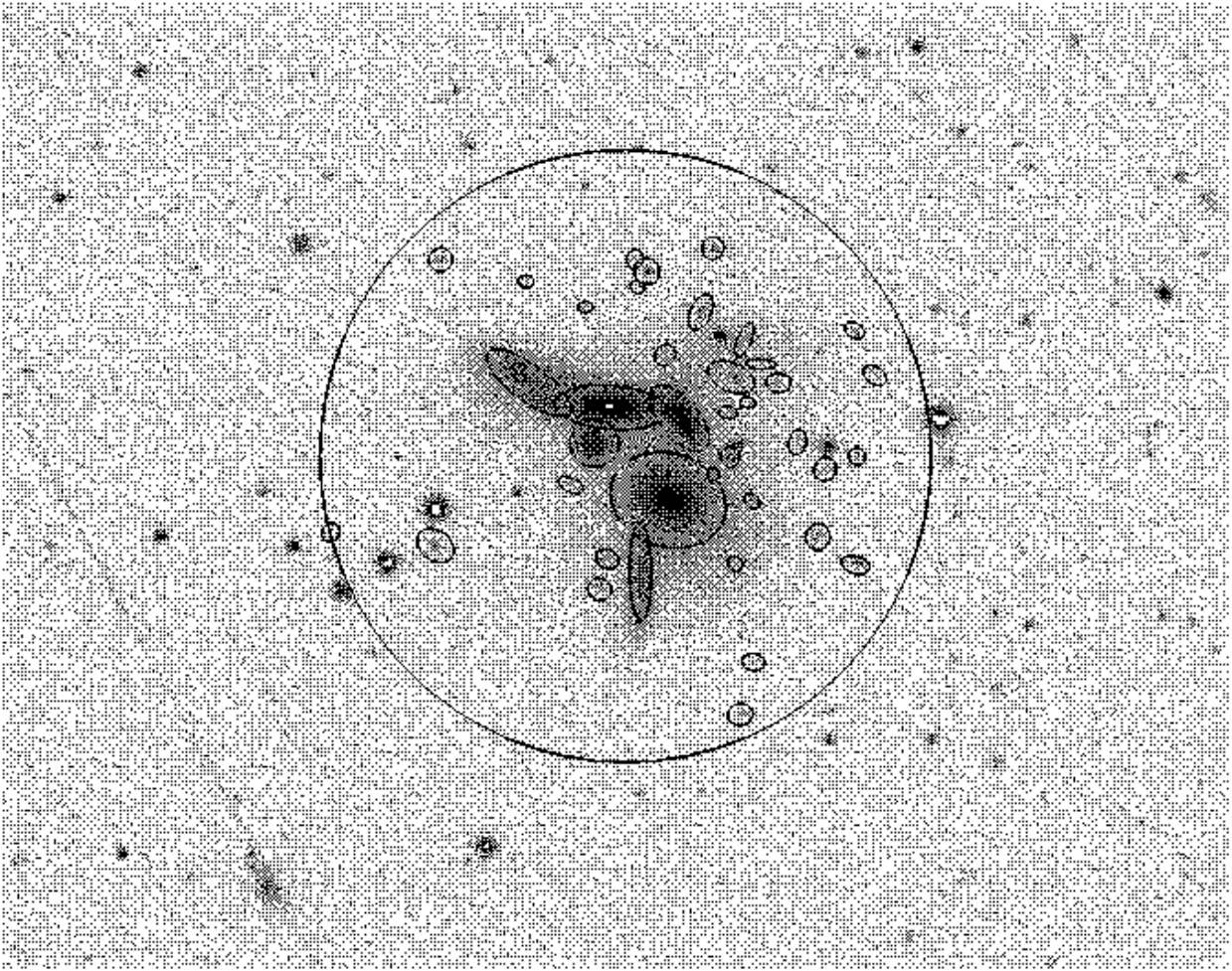}

\caption{Kitt Peak 0.9m image of Seyfert's Sextet.  Ellipses 
represent extended objects detected with the FOCAS software.
The large circle is constructed with a radius twice the size
of the radius of the smallest circle that just encloses the
23 mag arcsec$^{-2}$ isophotes of the giant galaxy members
of the Sextet.}

\label{kpnoimage}
\end{figure}

\citet{sdh98} have identified samples of dwarf galaxy candidates in a
number of HCGs using the FOCAS software package \citep{focas} to
perform star/galaxy separation on ground-based $R$ band images.
Seyfert's Sextet was not included in this study, however, we present
here (Figure \ref{kpnoimage}) a previously unpublished, deep $R$ band
image of Seyfert's Sextet observed after the conclusion of the HCG
survey of \citet{sdh98}.   Although the majority of the HCGs in the
\citet{sdh98} sample were observed with the Palomar 1.5m, this image
was taken with the Kitt Peak 0.9m telescope in 1996 May.  The data
reduction and analysis techniques used to process this image were
identical to those used in \citet{sdh98}.

In Figure \ref{kpnoimage}, a large circle indicates the boundary of the
group; the radius of this circle is twice the size of the radius of the
smallest circle that encloses the 23 mag arcsec$^{-2}$ isophotes of the
giant galaxy members, $2 \times R_g$ \citep[see][]{sdh98}, and within
that boundary every non-stellar object is indicated by an ellipse.
Apart from the objects that make up the Sextet (five galaxies and the
bright tidal tail), 33 objects were detected and classified as
non-stellar.  The majority of these objects are found in the northwest
quadrant of the image, and several appear to be superposed on the
fainter tidal tail, which extends to the northwest of NGC6027c.

\begin{figure}

\plotone{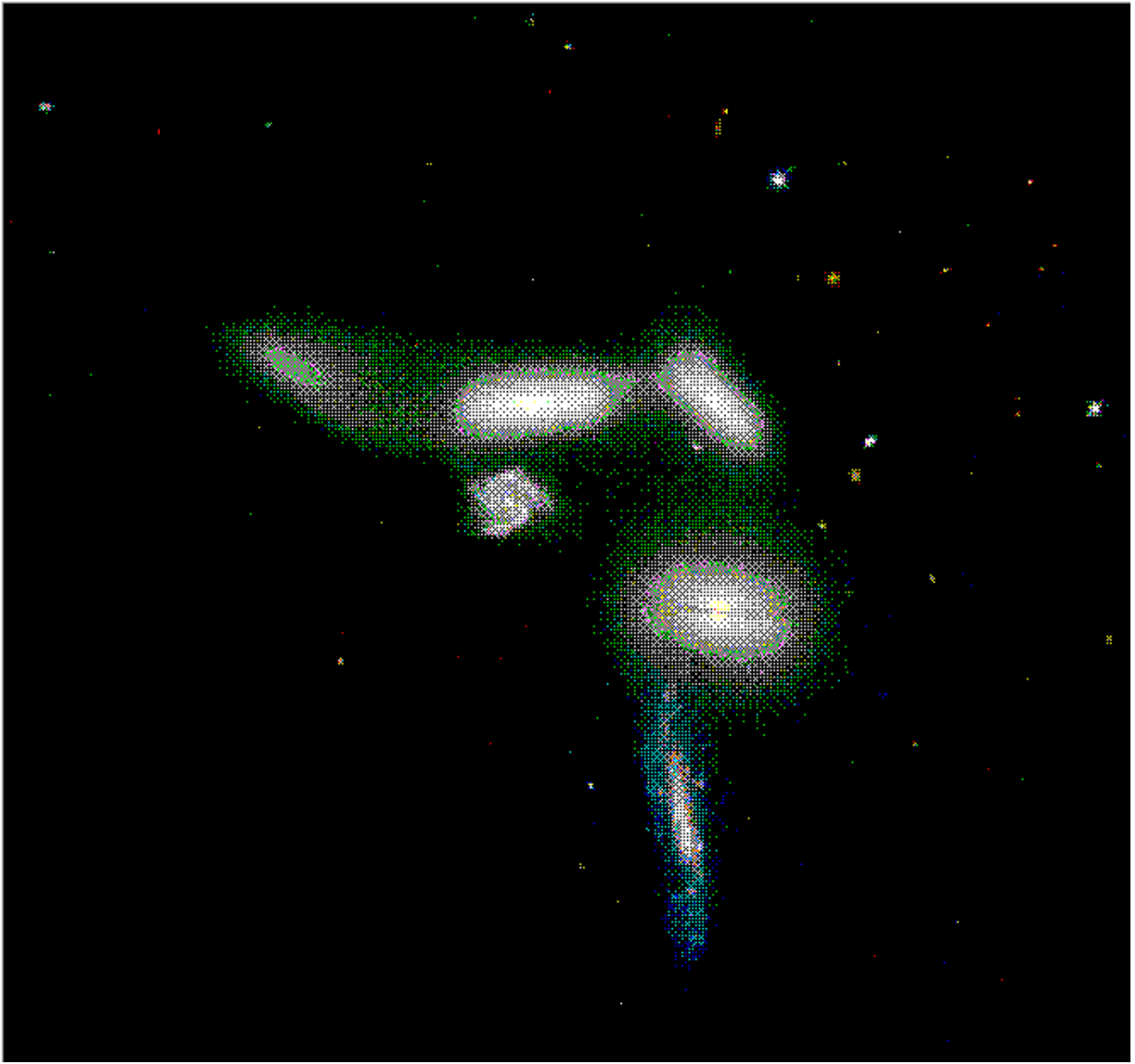}

\caption{A natural color image of Seyfert's Sextet created by combining
the black and white data taken through the four WFPC2 filters, F336W,
F439W, F555W, F814W ($UBVI$).  Colors assigned to these data
approximate the peak wavelengths of each filter: F336W data were
colored reddish-violet, F439W blue, F555W yellowish-green, and F814W
red.  The inclusion of the ultraviolet filter causes the disk of
NGC6027e, as well as the central regions of the edge-on spiral,
NGC6027d, to be rendered more pink than they would be if strictly
optical filters were used.}

\label{primage}
\end{figure}

\begin{figure}

\plotone{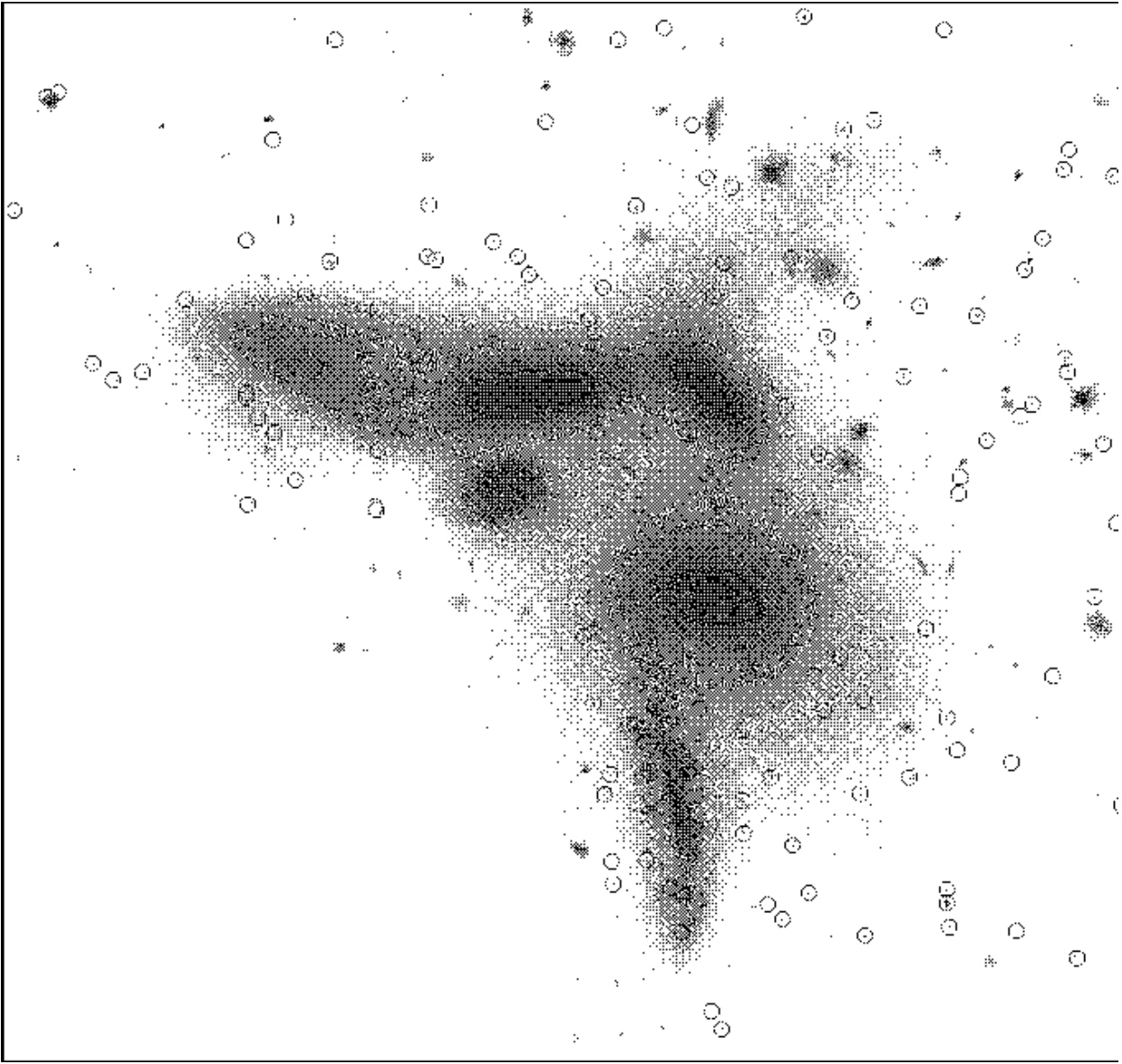}

\caption{A finding chart for the 188 star cluster candidates identified
on the three Wide Field chips.  This image is a logarithmically scaled
version of Figure \ref{primage}, which has been binned by a factor of 2
in $x$ and $y$ for image compression purposes.  A few sources are
partially or completely outside of the boundaries of this slightly cropped
image.  North is up, and east to the left.}

\label{fchart1}

\end{figure}

Comparison to the \textit{HST} image (Figure \ref{primage}) reveals
that the majority of the galaxies identified in the ground-based image
of the Sextet are background galaxies.  Viewed at higher angular
resolution, many of the non-stellar objects appear to be background
disk galaxies, and the two bright, red galaxies west of NGC6027c appear
to be background giant ellipticals.  However, several of the candidate
dwarf galaxies do appear to be just that, dwarf galaxies associated
with the Sextet.  Furthermore, several faint, diffuse objects not
visible in the ground-based image are also apparent in the \textit{HST}
image.  In this section we will describe the identification and
photometry of the extended sources observed in the \textit{HST}
image.

Detection of candidate dwarf galaxies associated with Seyfert's Sextet
is reasonably straightforward; assuming a distance modulus of 33.8, a
typical dwarf galaxy will have $M_V = -14$, or a total magnitude of
$\sim$20 in the F555W filter.  Furthermore, a dwarf galaxy with a core
radius of $\sim$300 pc will have an angular radius $>1\arcsec$, and
will therefore cover roughly 300 or so pixels.  Thus, we expect most
dwarf galaxies (including those fainter than $M_V = -14$) to be
detected with high signal-to-noise in our F555W and F814W images.
There are some limits, though; the photometric error is $\sim0.1$ for
point sources with $M_V < -9$, and the sensitivity of the WF chips to
low surface brightness sources is limited.  Thus, we do not expect to
detect the faintest, diffuse dwarf galaxies, similar to the Local Group
dwarf spheroidal galaxies (which are as faint as $M_V \sim -8$).

Selection of extended sources to photometer was accomplished using the
FOCAS software package, however, we set the parameters in order to
select relatively bright, extended sources.  The catalogue was
constructed to include all objects with a minimum area of 50 pixels
(0.5 square arcseconds) with fluxes 2.5$\sigma$ above the background.
This catalogue did contain some obvious contaminants, including
saturated stars and several resolved \ion{H}{2} regions in NGC6027d and
NGC6027e.  These were removed by hand.  Ultimately, we were left with
38 objects.  Most of the objects identified in the ground-based image
that are within the boundaries of the WFPC2 image were included, as
well as a number of fainter objects not visible in Figure
\ref{kpnoimage}.  The only objects identified in the ground-based image
that were not recovered by the automated detection algorithm were those
found very near the boundaries of the WF chips, where the images had
been trimmed.  We also note than in a few cases, objects identified
from the ground as a single extended source were resolved into two
extended sources in the \textit{HST} images.

It is apparent by visual inspection that a number of the extended
sources seen in Figure \ref{primage} are disk galaxies.  We remove from
further consideration as potential dwarf members of Seyfert's Sextet
those galaxies with clear spiral structure and those that appear to be
edge-on spiral galaxies.  We assume that these objects are background
galaxies.  For the final catalogue of candidate dwarf galaxies, we have
photometered 29 objects; six from the WF2 chip, 17 from WF3, and six
from WF4.  In addition, we have photometered six of the background
spiral galaxies (the other three were found on the edges of two WF
chips, in the region trimmed from the images during data reduction) for
comparison to the sample of dwarf galaxy candidates.  Figure
\ref{fchart2} is a finding chart for the 29 dwarf galaxy candidates and
six background spiral galaxies that were photometered; a contour plot
of a smoothed version of Figure \ref{primage} (smoothed to reduce some
of the noise in the fainter isophotes) is shown with square boxes
marking each object in this sample.

For each extended object in the catalogue, the STSDAS task ELLIPSE was
used to fit elliptical isophotes to the galaxy in the \textit{HST} $I$
image out to a semi-major axis where the signal to noise became too low
for the fitting routine to work successfully.  The ellipse parameters
derived for each object from the $I$ image were then used to calculate
the object's flux in the other filters.  A value of the average
background flux per pixel was calculated for each image.  For each
object, the total flux within the elliptical apertures was corrected by
subtracting off the total background contribution within the aperture,
assuming that this average background flux per pixel is flat across the
image.  After background subtraction, the total flux for each object
was converted into a total magnitude using the adopted
\citet{dolphin00} zero points \citep[which were corrected to infinite
aperture using the standard 0.1 magnitude offset;][]{holtzman95}.  Few
objects were detected in either $B$ or $U$, and those that were
detected have low $S/N$.  Figure \ref{extCMD} is a $V - I$ vs. $V$
color magnitude diagram for the 29 extended sources in the final
catalogue.  All of the photometry has been corrected for Galactic
reddening.  For reference, the point sources from Figure \ref{CMD} are
plotted as well.

\subsection{Results of the Extended Source Photometry}

\begin{figure}

\plotone{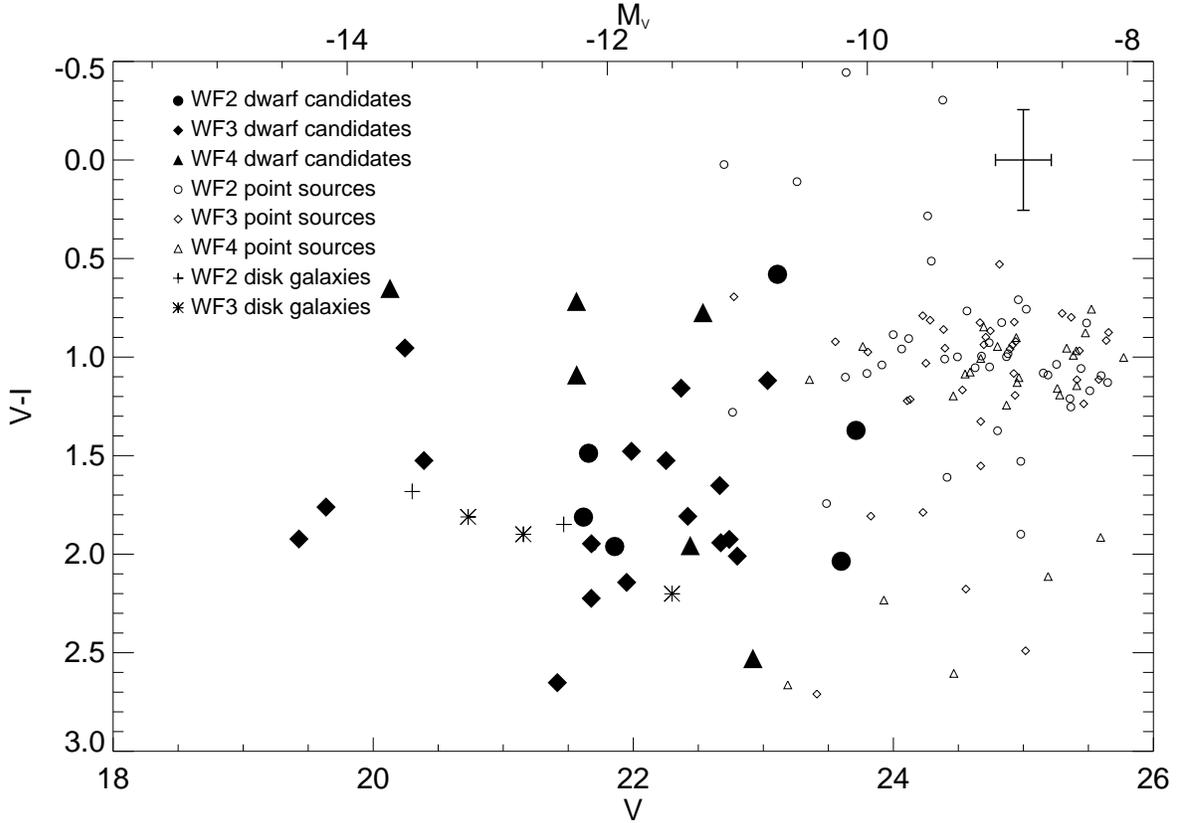}

\caption{Color-magnitude diagram for extended sources detected
in $V$ (F555W) and $I$ (F814W).  The filled symbols represent
extended sources on WF2 (filled circles), WF3 (filled diamonds),
and WF4 (filled triangles).  For reference, the point sources
from Figure \ref{CMD} are included as open polygons.  The error
bars in the upper right represent the mean error in the photometry
of all extended objects.  All of the photometry has been
corrected for Galactic reddening.  The absolute magnitude scale
(top) is given for the objects at the distance of the Sextet,
57.5 $h_{75}^{-1}$ Mpc.}

\label{extCMD}
\end{figure}

As noted previously, when viewed at higher angular resolution, many of
the extended objects in the field of Seyfert's Sextet seen in Figure
\ref{kpnoimage} appear to be background galaxies rather than dwarf
galaxies associated with this HCG.  The photometric results for the
extended sources (Figure \ref{extCMD}) lend some credence to this
supposition.  The extended objects are not distributed randomly around
the group, instead they appear concentrated to the northwest of the
Sextet.  While we lack information on the southeast quadrant for
comparison (even if the PC data was analyzed, it covers significantly
less area than the WF chips), the density of extended objects on WF3 is
noticeably higher than on WF2 or WF4.  As expected for a background
cluster or group, the majority of the galaxies observed lie in a narrow
region of parameter space, with $V - I > 1.5$ and $21 < V < 23$.  The
six extended sources that are obvious disk galaxies are also found in
this same region of the color-magnitude diagram.  In addition to the
faint, red galaxies, there are three bright, red objects on WF3 ($V - I
> 1.5$ and $V < 20.6$) that appear to be the massive elliptical galaxy
members of a poor cluster or group.

Based on color alone it is difficult to say with certainty that the
three bright ellipticals are giants and not dwarfs because
the $V - I$ colors of dwarf galaxies and giant ellipticals can be
similar.  For dwarf galaxies, $V - I$ colors of $\sim 1 - 1.3$ are
typical.  For example, the Local Group dwarf ellipticals (dEs) NGC
147, NGC 185, and NGC 205 have $V - I = 1.12, 1.25, 1.23$ respectively
\citep{mateo98}.  In a recent study of low surface brightness dwarf
galaxies in the Dorado Group, \citet{carrasco01} found 65 such objects
with  $V-I$ colors from $-0.3$ to 2.3.  This distribution had a strong
peak at $V - I = 0.98$, again suggesting that this is a typical
color for a dE galaxy.  For giant elliptical galaxies (gEs), the
typical $V - I$ color on average may be redder by a few tenths of a
magnitude than that of a typical dE.  For example, \citet{carollo97}
observed a sample of 15 nearby ellipticals with \textit{HST} and found
a range of colors of $1.24 < V - I < 1.42$.   Although the typical
colors of gEs and dEs may only differ by a small amount, there is overlap
in the overall distribution of colors between these two populations
of galaxies.  Thus, additional information besides the broadband
colors is required to distinguish between a dE and a gE.

\begin{figure}

\plotone{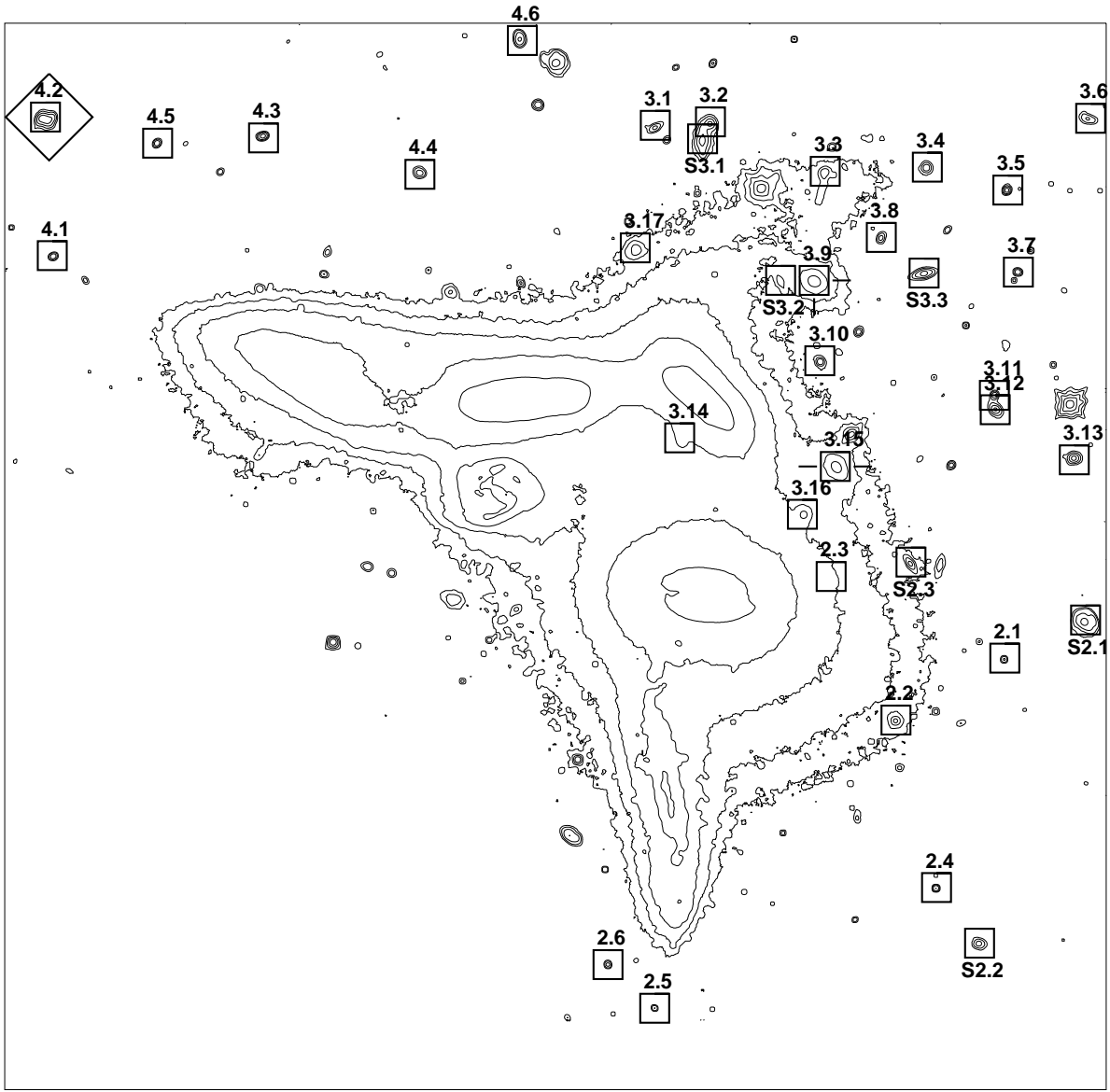}

\caption{Finding chart for the 29 candidate dwarf galaxies and seven
background spiral galaxies that were photometered.  The image displayed
in Figure \ref{primage} was smoothed with a 3 pixel boxcar averaging
function to reduce some of the noise in the fainter isophotes and then
contoured.  Each galaxy that was photometered is marked with a large
square and is labeled with an identifying number.   For the dwarf
galaxy candidates, the identifier is of the form X.X, where the first
number is the WF chip and the second number is a sequence number.   The
background spirals are given identifiers of the form SX.X.  A redshift
has recently been obtained for Galaxy 4.2 (indicated by the large, open
diamond) with the Hobby--Eberly Telescope; see \S3.  Surface brightness
profiles for the two bright, red ellipticals (objects 3.9 and 3.15,
indicated by two dashes) are presented in Figure \ref{surfprof}.}

\label{fchart2}

\end{figure}

The majority of the galaxies identified in Figure \ref{fchart2} have
colors redder than either the typical dE or gE.  The amount of Galactic
reddening towards Seyfert's Sextet is low, and has already been removed
from the colors plotted in Figure \ref{extCMD}.  Internal, differential
reddening due to dust in the Sextet system is likely to affect the
colors of the galaxies to some extent, but $A_{V}$ of $2-3$ magnitudes
is required to redden dwarf galaxies with $V - I \sim 1.0$ to $V - I >
1.5 - 2.0$.  Since the majority of the galaxies photometered are
located outside of the bright isophotes of the Sextet galaxies, it is
unlikely that they are being reddened to such a significant extent.

The assumption that the majority of the faint, red galaxies observed
(including the six background spiral galaxies, which have red $V - I$
colors) are members of a background cluster provides a convenient
explanation for the red colors of these galaxies without invoking
significant internal reddening.  Assuming that the three bright
ellipticals are giant ellipticals, then the redshift of this group is
likely to be $z > 0.3$ so that $M_V < -20.5$ for these objects.  At $z
= 0.32$, the $k$ correction to the $V - I$ color is $\sim$0.67
magnitudes for an elliptical galaxy and $\sim$0.40 magnitudes for an SA
galaxy \citep{kcor}.  Thus, we expect an elliptical at $z \sim 0.3$ to
have $V - I \sim 1.9 - 2.1$, which is similar to the colors of the
ellipticals found in our sample.

\begin{figure}

\epsscale{0.67}
\plotone{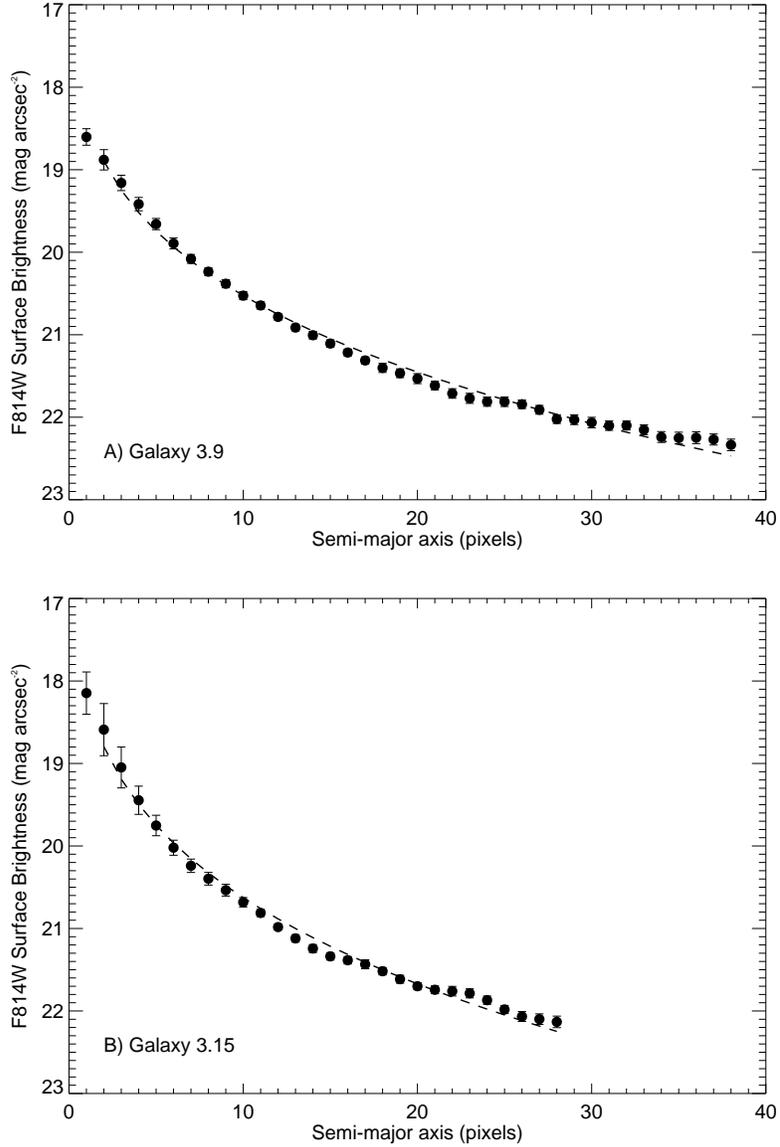}

\caption{F814W surface brightness profiles for galaxies 3.9 (top panel)
and 3.15 (bottom panel).  Overplotted on the data are fits to a
modified S\'{e}rsic profile with shape parameter $n=1/4$, which is
equivalent to the de Vaucouleurs $r^{1/4}$ law.  Both objects are fit
well with this type of profile and have central surface brightnesses of
$15.6$ (galaxy 3.9) and $15.1$ (galaxy 3.15) mag arcsec$^{-2}$.  The
shapes of the profiles and the central surface brightnesses are
consistent with those of giant elliptical galaxies.}

\label{surfprof}

\end{figure}

Perhaps the strongest evidence that these three bright, red objects are
not dwarf galaxies but are instead giant galaxies comes from their
surface brightness profiles.  Figure \ref{surfprof} presents $I$ band
surface brightness profiles for the two brightest ellipticals, which
are labeled objects 3.9 and 3.15 on the finding chart (Figure
\ref{fchart2}).  The third bright elliptical (galaxy 3.16) is fainter
than galaxies 3.9 and 3.15 by more than a magnitude, and its profile
was not fit because the lower signal to noise surface photometry
resulted in a poorly constrained fit.  The two brightest elliptical
galaxies are fit well by S\'{e}rsic profiles of the form:

\begin{equation}
\mu(r) = \mu_{0} + 1.0856(r/r_{0})^n
\end{equation}

\noindent where $\mu$ is the F814W surface brightness in mag
arcsec$^{-2}$ and $n$ is fixed at $1/4$ (i.e., the de Vaucouleurs law).
In Figure \ref{surfprof}, these $r^{1/4}$ law fits are overplotted
on the surface brightness data, showing the goodness of fit.  For
galaxy 3.9, the fitted central surface brightness, $\mu_{0}$, is $15.6
\pm 1.0$ mag arcsec$^{-2}$ and for galaxy 3.15, $\mu_{0} = 15.1 \pm
1.2$ mag arcsec$^{-2}$.  Recent studies of dwarf galaxies in Virgo
\citep{durrell97, bin98} have shown that $r^{1/4}$ law fits are
inappropriate for dE galaxies; instead, the ``shape parameter'', $n$,
is allowed to vary, and dEs show a correlation between total magnitude
and $n$.  The range in $n$ in dEs is found to be $0.4 < n < 2.0$ in
both studies.  Thus, the goodness of fit of an $r^{1/4}$ law to
galaxies 3.9 and 3.15 is evidence that they are unlikely to be dEs.
Furthermore, the central surface brightnesses of Virgo dEs ranges from
17.8 to 24.9 in $R$ \citep{durrell97} and from 18.1 to 25.3 in $B$
\citep{bin98}.  We conclude that both the shapes and central surface
brightnesses of galaxies 3.9 and 3.15 are consistent with their being
background gE galaxies rather than dEs associated with Seyfert's
Sextet.

Among the sample of extended objects, there are eight galaxies with $V
- I$ colors less than 1.2, similar to the typical values for known dEs,
and thus these objects are considered candidate dwarf galaxies.  Table
\ref{dwcand} includes the names and photometric results for these eight
objects.   Among these eight galaxies, six are simply faint ($V
\lesssim 22$), blue ($V - I < 1.2$) galaxies.  One of these six (galaxy
2.6) lies near the edge of the disk of NGC6027d, while two others
(galaxies 4.1 and 4.3) lie near the bright tidal tail associated with
NGC6027b.  Galaxy 3.8 is a faint, blue galaxy that is found among the
galaxies that form the suspected background cluster or group on WF3.
However, it is found just outside of the isophotes of the tidal tail
that extends from NGC 6027c.  Galaxy 3.10 is also found among the
suspected background cluster, but is further from the tidal feature.
Galaxy 4.6 may be a background spiral, but the spiral structure is not
clear enough to definitively classify this object as a disk galaxy.

Besides the six faint, blue dwarf galaxy candidates there are two
objects with more irregular morphologies.  Galaxy 4.2 is a bright, blue
galaxy that shows obvious signs of star formation in the form of
several compact, bright point sources superposed on a faint,
\textsf{S}-shaped continuum.  This object's morphology is reminiscent
of the class of blue compact dwarfs \citep[see, e.g.,][]{papa96}.
Galaxy 3.14 is an irregularly shaped, extended object seen quite close
to the disk of NGC6027c.

In addition to the color selected candidates, we include one other
galaxy in Table \ref{dwcand} as a possible dwarf galaxy candidate.
Galaxy 3.3 is one of the ``faint, red'' galaxies found on chip WF3.
However, this galaxy is found within the boundaries of the faint tidal
tail associated with NGC6027c, and it has a very peculiar morphology.
This object appears to have a ``cometary'' morphology, with an
elliptical galaxy at the ``head'' and a narrow ``tail'' of emission
extending to the south of the head.  This galaxy appears to be
a single, distorted galaxy, however, we note that a more nearby
pair of galaxies exhibit a similar appearance.  The system
Arp 296 \citep[see][for a $B$ band image]{hibbardyun} is a face-on
disk galaxy and a thin, edge-on disk galaxy that may be similar
to galaxy 3.3.  If galaxy 3.3 is a system like Arp 296, then it is
most likely a background object rather than a dwarf galaxy associated
with the Sextet.

These nine candidate dwarf galaxies stand out among the sample of 
extended objects because of their colors, and also, in the three
specific cases described previously, because of their morphologies.
It is interesting that many of these faint galaxies lie in or near
the tidal features associated with the Sextet, however, this may be
coincidental since the Sextet galaxies fill much of the area of the
three WF chips.  In the end, we refer to these galaxies as candidate
dwarfs, since redshift information is necessary to discriminate between
the possibilities that they are dwarf galaxies in the Sextet or are
background galaxies.

We hypothesized that the majority of the faint, red galaxies observed
are members of a group or cluster seen in projection against Seyfert's
Sextet.  The observations indicate that the bright, red elliptical
galaxies are background giant ellipticals rather than dwarf
ellipticals.  However, there is little direct evidence that all of the
faint, red galaxies are also background objects; certainly the colors 
(assuming a relatively significant $k$ correction) and morphologies of
the objects \textit{support} the hypothesis that these objects are
background objects, but they do not prove the hypothesis.  It is possible
that some of the red galaxies may be dwarf galaxy members of Seyfert's
Sextet, but it does seem more likely that the majority are background
galaxies.

\section{Spectroscopy of a Tidal Dwarf Candidate}

During 2002 March, spectroscopic follow-up observations of a number of
galaxies identified in the \textit{HST} images of Seyfert's Sextet were
undertaken with the Hobby-Eberly Telescope \citep{het}.  The
observations were made with the Marcario Low-Resolution Spectrograph 
\citep{hetlrs, schneiderlrs} in multi-object mode.  Three 800 second
exposures were obtained and subsequently averaged together.
Unfortunately, observing conditions were such that only one spectrum
(galaxy 4.2) was extracted with reasonable signal-to-noise.

\begin{figure}

\epsscale{1.00}
\plotone{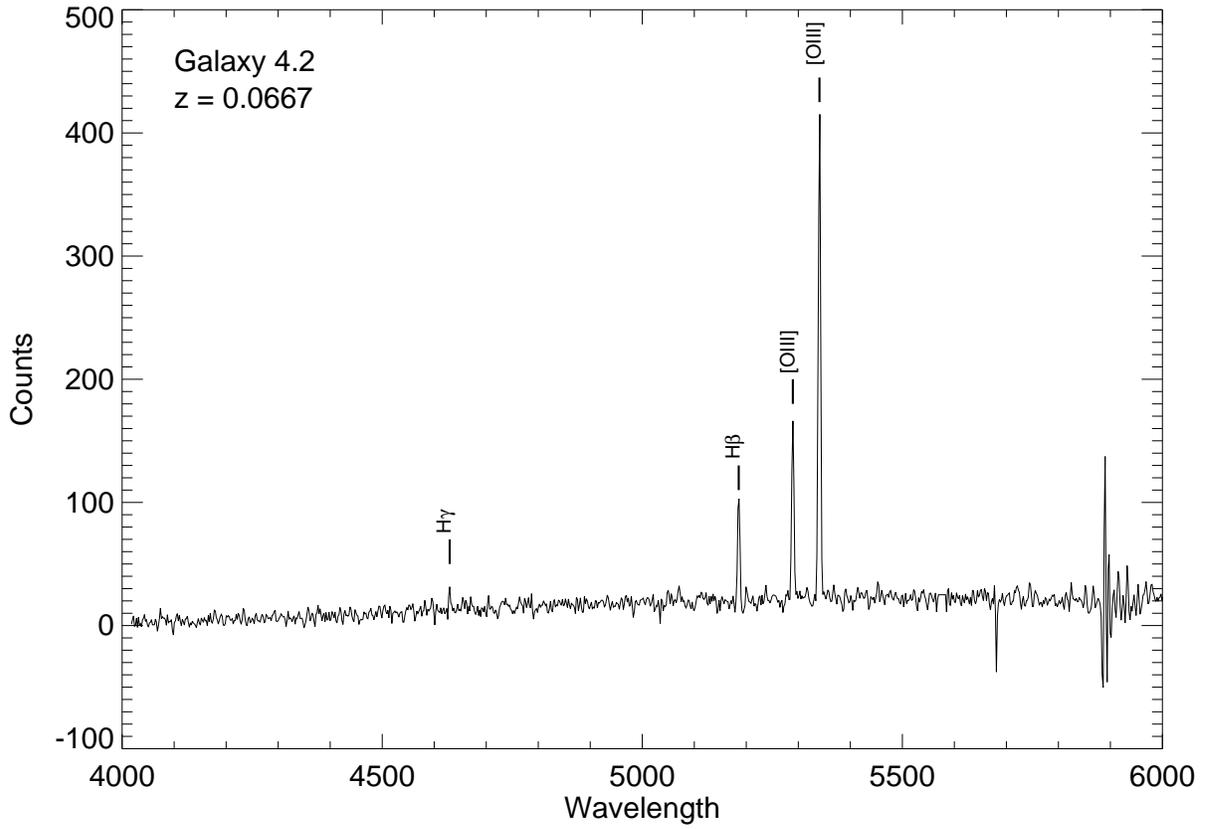}

\caption{Hobby-Eberly Telescope spectrum of galaxy 4.2.  Four emission
lines are labeled, H$\gamma$, H$\beta$, and the \ion{O}{3} $\lambda\lambda$
4959, 5007 doublet.  The redshift for this galaxy is $z = 0.0667$, which
is nearly identical to that of the discordant spiral galaxy found
in Seyfert's Sextet, NGC6027e.}

\label{spectrum}

\end{figure}

The observations were obtained with a 600 line mm$^{-1}$ grism used in
conjunction with $1\farcs3$ slitlets to provide resolution of $R \sim
1000$.  Wavelength calibration for the spectrum was determined from
both HgCdZn and Neon comparison lamp exposures; the resulting
wavelength calibrated spectrum covers $\sim 4000 - 6900$ \AA.
Figure \ref{spectrum} shows the wavelength calibrated spectrum of galaxy
4.2 over the range $4000 - 6000$ \AA.

Although a weak continuum is detected, the spectrum does contain four
emission lines; H$\beta$ and the $[$\ion{O}{3}$]$ $\lambda\lambda$4959, 5007
doublet are observed at high signal-to-noise, while H$\gamma$ is detected
with marginal significance.  These lines and the observed wavelengths for each
(derived from Gaussian fits to the line profiles) are listed in Table 
\ref{redshifts}.

The mean redshift for the accordant galaxies in Seyfert's Sextet is $v
= cz = 4347$ km s$^{-1}$ \citep{hickson92}.  The emission lines provide
a redshift for this dwarf galaxy candidate of $v = cz \sim$20,000 km
s$^{-1}$, placing it in the background behind Seyfert's Sextet.
However, it is interesting to note that this galaxy has a redshift
almost identical to that of NGC6027e, the discordant spiral in the
Sextet, which has a redshift of 19,809 km s$^{-1}$ \citep{hickson92}.
Adopting $z=0.067$ for NGC 6027e and galaxy 4.2, the distance to these
two galaxies is approximately 260 $h_{75}^{-1}$ Mpc.  The absolute $V$
and $I$ magnitudes of galaxy 4.2 are approximately $-16.7$ and $-17.4$
respectively.  The projected separation between these two galaxies is
roughly 80\arcsec, which at the adopted distance is $< 100$ kpc.  Thus,
we suggest that galaxy 4.2 may be a dwarf galaxy satellite of
NGC6027e.

\section{Discussion and Conclusions}

The case study of Seyfert's Sextet presented here is part of a
continuing effort to determine whether dwarf galaxies form during tidal
interactions among giant galaxies.  The Sextet appears to be the most
logical choice to search for tidal dwarf formation; it is the most
compact of the Hickson compact groups, contains two prominent tidal
tails, has a low velocity dispersion, and previous ground-based imaging
revealed a number of faint, extended objects within the boundaries of
the group.  However, the results of the \textit{HST} imaging of
Seyfert's Sextet show that, contrary to expectations, there is very
little evidence for dwarf galaxy formation or any other strong star
formation in this group.

A large number of both point sources and extended sources were 
catalogued and photometered from the three Wide Field images.
We find that very few objects are detected in either of the
two blue filters, F336W and F439W, and those that are detected
in the two red filters, F555W and F814W, have red colors consistent
with those of old stellar populations.  The majority of the
point sources detected appear to be old ($> 1$ Gyr) and the majority
of the extended sources detected appear to be background galaxies.

These photometric results contrast sharply with \textit{HST} imaging
studies of other HCGs, such as HCG92 (Stephan's Quintet) and HCG31.  In
HCG92, \citet{sq01} found a number of bright, blue star cluster
candidates in the tidal debris regions of this group.  The images of
this compact group also show bright, blue extended sources in the
``Northern Starburst Region'' and in the tidal tails of NGC7319 and
NGC7318a/b.  HCG31 contains a significant number of bright, blue point
sources \citep{johnson00} similar to those seen in HCG92.  Star-forming
regions are also observed in HCG31 that are ``too small to be called
galaxies themselves, but are not clearly associated with either galaxy
AC or galaxy E'' \citep{johnson00}.  Thus, both HCG92 and HCG31 contain
what appear to be young star clusters and tidal dwarf galaxy candidates,
while Seyfert's Sextet does not appear to contain a significant population
of either type of object.

The star cluster candidates identified in this study have photometric
properties consistent with those for models of massive ($\sim10^{6}
M_{\sun}$) clusters with ages $10^{8.5} - 10^{9.5}$ years.  The ages
of these objects suggest that they are not entirely a primordial
population, but may be the product of an interaction within the compact
group at some time within the past few Gyr.  \citet{williams91} argue 
that the optical tail associated with NGC 6027b and the \ion{H}{1}
gas that they associate with NGC 6027d may have resulted from an interaction
between these two disk galaxies more than $5 \times 10^{8}$ years ago.
The ages we derive for many of the cluster candidates are consistent
with this hypothesis.

While there do not appear to be any young star clusters or tidal dwarf
galaxies associated with Seyfert's Sextet, we did identify several
candidate dwarf galaxies in the group.  This sample includes a few
faint, blue extended sources, and two galaxies with peculiar
morphologies:  an irregularly shaped galaxy located quite near the disk
of NGC6027c and an unusual, ``cometary'' galaxy located within the
tidal tail associated with NGC6027c.  An additional candidate, galaxy
4.2, has already been ruled out as a member of the Sextet; the
Hobby-Eberly Telescope spectrum of this object instead shows that it is
associated with NGC6027e, the discordant redshift member of Seyfert's
Sextet.  Whether or not the other candidate dwarf galaxies are
associated with the Sextet, they appear morphologically very different
from the clumpy, blue tidal dwarf galaxy candidates in HCG92 and
HCG31.  

The data suggest that there is some fundamental, physical difference
between the Sextet and the two HCGs that are known to contain young
star clusters and tidal dwarf galaxy candidates.   One obvious
difference between these groups are the types of galaxies contained in
each:  Seyfert's Sextet is primarily made up of early-type (S0/E)
galaxies, Stephan's Quintet contains spirals, and HCG31 contains mostly
irregular galaxies.  Based on these morphologies, one initial
expectation is that the neutral gas content in Seyfert's Sextet is
likely to be lower than that of either HCG31 or HCG92.  Radio
observations show that the Sextet contains only $2 \times 10^{9}\:M_{\sun}$ 
of neutral Hydrogen \citep{vm01}, about an order of magnitude
less than that of HCG31 and HCG92 \citep{williams91, vm01}.  The most
recent observations of the gas content of HCG92 \citep{williams02}
revise the gas mass of this group downward, however it remains at least
five times larger than the gas in Seyfert's Sextet.

What appears to be the more significant difference among these three
HCGs, however, isn't the gas mass, but the \textit{distribution} of the
\ion{H}{1}.  \citet{williams91} present VLA neutral Hydrogen observations
of Seyfert's Sextet that indicate that the majority of
the \ion{H}{1} mass is retained by the disk of NGC6027d, although some
gas is found in a tail extending to the east of this galaxy and also in
the optical tidal tail associated with NGC6027b.  In HCG31, the VLA
neutral Hydrogen maps \citep{williams91} show that the gas is found
both in the galaxies themselves and in a large envelope of gas that is
plausibly attributed to tidal interactions between the galaxies.  The
distribution of \ion{H}{1} in HCG92 is found to lie entirely outside of
the galaxies \citep{vm01, williams02}, however.  The \ion{H}{1} is
concentrated in clouds and tidal tails that are not coincident with the
disks of the member galaxies.

\citet{vm01} proposed an evolutionary sequence based on their VLA
observations of the \ion{H}{1} content of HCGs.  In their model,
``phase 1'' HCGs are those where the vast majority of the neutral gas
remains bound to the member galaxies.  ``Phase 2'' HCGs are more
evolved in the sense that the galaxies retain some of the gas, while
approximately half of the gas mass is found in tidal features.  The
final, most evolved phase is broken into two subclasses, ``phase 3a''
and ``phase 3b''.  Phase 3a groups are those where the gas is almost
completely stripped from the galaxies and is found entirely within
tidal features, while phase 3b groups are a few rare cases where the
entire group seems to be contained in a single \ion{H}{1} cloud.  HCG31
is considered a prototype phase 2 group, and HCG92 is considered an
extreme example of phase 3a.  Both \citet{williams91} and \citet{vm01}
find Seyfert's Sextet to be anomalous; its gas distribution suggests
that the system has not experienced significant dynamical evolution,
while optical observations suggest the opposite.

We propose one possible scenario for the history of Seyfert's Sextet
that takes into account the following significant factors:  (1)  The
tidal tails are evidence for interactions among the accordant redshift
members some time in the past, (2) the interactions that have occurred
have not triggered star and/or star cluster formation similar to that
seen in other merging galaxies and compact groups, (3) the relatively
small amounts of neutral gas in Seyfert's Sextet remains bound in the
one late type galaxy and does not appear to be distributed among the
group environment, and (4) the low velocity dispersion among the member
galaxies and the small distances between the member galaxies suggests
that future interactions among the galaxies are likely.  This
accumulated evidence suggests that a number of gas-poor (and one
gas-rich) galaxies have interacted beginning perhaps as long as 1 Gyr
or more in the past (dated by the colors of the red globular cluster
candidates).  The interactions in the group have created the optical
tidal tails and perhaps created the elliptical member of the Sextet,
NGC6027a, as well.  The interactions in the past stripped stars from
the progenitor galaxies, redistributing them within the group.  The
evidence for a red, low surface brightness halo encompassing all of the
member galaxies, which is seen in our images as well as deeper
ground-based images, is further evidence for a redistribution of the
galaxies' stars within the group.  The only ongoing star formation and
most of the neutral gas is found within the disk of NGC6027d, the only
late type member of the group, suggesting that any interaction that
involved this galaxy must have been minor, although the galaxy disk
does appear somewhat irregular and perhaps warped.

We speculate that further interactions are probably inevitable, and a
major interaction between NGC6027d and the other members of the group
may trigger the stripping of its neutral gas and star cluster formation
throughout the group in the future.  Moreover, the low velocity
dispersion suggests that none of the four large galaxies are likely to
escape the group, and thus the group members may merge into a single
galaxy, rather than remaining distinct.  Thus, we believe that we are
seeing Seyfert's Sextet at the ``beginning of the end''; we presume
that the future interactions will be the end of this group,
transforming it into a single galaxy.

\acknowledgements

We wish to acknowledge the help of M. Eracleous with data reduction of
our HET spectra.   We also wish to thank J. Hibbard for a number of
useful discussions.  SZ acknowledges support from the National Science
Foundation through a Research Experiences for Undergraduates award.
This work was supported by NASA STScI and by the NSF under grants NSF
AST 00-71223 and STSI HST-GO-08717.04-A.

\clearpage

\clearpage

\begin{deluxetable}{lllcc}
\tablewidth{450pt}
\tablecaption{Properties of Seyfert's Sextet}
\tablehead{\colhead{Property\tablenotemark{a}}}
\startdata
Alternate Names: & & HCG 79, VV115, NGC6027 \\
Equatorial Coordinates: & & 15$^{\rm h}$59$^{\rm m}$12$^{\rm s}$ $\:\:$
                                  $+20\arcdeg45\arcmin31\arcsec$ (J2000.0)\\
Mean Redshift: & & 0.0145 \\
Radial Velocity Dispersion: & & 138 km s$^{-1}$ \\
\\
Galaxies:&  & Name & Type   &  $v$ (km s$^{-1}$) \\ \cline{3-5}
         & & NGC6027a & E0 &  \phn4294 \\
         & & NGC6027b & S0 &  \phn4446 \\
         & & NGC6027c & S0 &  \phn4146 \\
         & & NGC6027d & Sdm       &  \phn4503 \\
         & & NGC6027e & Scd       &  19809 \\
\enddata

\tablenotetext{a}{Data taken from \citet{hicksonbook}.}

\label{SSparams}

\end{deluxetable}

\begin{deluxetable}{lcrr}
\tablewidth{175pt}
\tablecaption{Aperture Corrections}
\tablehead{\colhead{Filter} & \colhead{$N_{stars}$} & \colhead{Value} & 
\colhead{$\sigma$}}
\startdata
F814W & 17 & $-0.199$ & 0.022 \\
F555W & \phn8 & $-0.149$ & 0.033 \\
F439W & \phn6 & $-0.163$ & 0.025 \\
F336W & \phn6 & $-0.144$ & 0.021 \\
\enddata
\label{apcor}
\end{deluxetable}

\begin{deluxetable}{lrrrr}
\tablewidth{300pt}
\tablecaption{Photometric results for dwarf galaxy candidates}
\tablehead{\colhead{ID} & \colhead{F814W\tablenotemark{a}} & 
\colhead{F555W\tablenotemark{a}} & 
\colhead{F439W} & \colhead{F336W}}
\startdata
2.6\tablenotemark{b} & 22.63$\pm$0.25 & 23.28$\pm$0.19 & $\sim 25.2$ & 
$\sim 24.0$ \\
3.3\tablenotemark{c} & 20.83$\pm$0.11 & 22.42$\pm$0.19 & $\cdots$ & $\cdots$ \\ 
3.8   & 22.02$\pm$0.20 & 23.20$\pm$0.25 & $\cdots$ & $\cdots$ \\
3.10  & 21.31$\pm$0.18 & 22.54$\pm$0.23 & $\cdots$ & $\cdots$ \\
3.14  & 19.39$\pm$0.06 & 20.42$\pm$0.06 & 21.39$\pm$0.28 & 21.70$\pm$0.59 \\
4.1\tablenotemark{b}   & 21.86$\pm$0.22 & 22.71$\pm$0.20 & $\sim 24.6$ & 
$\sim 23.4$ \\
4.2   & 19.58$\pm$0.15 & 20.30$\pm$0.13 & 21.37$\pm$0.22 & 20.98$\pm$0.42 \\
4.3   & 20.95$\pm$0.12 & 21.73$\pm$0.13 & 23.01$\pm$0.27 & 22.21$\pm$0.35 \\
4.6   & 20.58$\pm$0.12 & 21.74$\pm$0.15 & $\cdots$ & $\cdots$ \\
\enddata

\tablenotetext{a}{The magnitudes listed here have not been corrected for Galactic
extinction.  The values adopted for the extinction/reddening corrections are given 
in \S2.3.}

\tablenotetext{b}{For these two cases, the objects were detected in the two blue
filters, but at low signal to noise.  Rough limits to the
magnitudes are provided, but the errors are likely to be $>0.5$ magnitudes.}

\tablenotetext{c}{This object is one of the ``faint, red'' galaxies on WF3,
but it is included here due to its peculiar morphology and presence in the
tidal debris of NGC6027c.  The photometry presented here only represents the
``head'' of the galaxy and does not include the ``tail'' of emission 
that appears to be associated with this object.}

\label{dwcand}
\end{deluxetable}

\begin{deluxetable}{lcc}
\tablewidth{230pt}
\tablecaption{Emission Lines Detected in Spectrum of Galaxy 4.2}
\tablehead{\colhead{Line} & \colhead{$\lambda_{obs}$ (\AA)}  &
\colhead{redshift (km s$^{-1})$\tablenotemark{a}}}
\startdata
H$\gamma$ & 4629.5 & 19,960 \\
H$\beta$ & 5185.4 & 19,987 \\
$[$\ion{O}{3}$]$ $\lambda$4959 & 5289.5 & 19,984 \\
$[$\ion{O}{3}$]$ $\lambda$5007 & 5340.7 & 19,988 \\
\enddata

\tablenotetext{a}{For consistency with the heliocentric velocities of
HCG galaxies reported in \citet{hickson92}, the redshift values
listed here were calculated as $v = cz$.}

\label{redshifts}
\end{deluxetable}

\end{document}